\documentclass[lettersize,journal]{IEEEtran}
\usepackage{amsmath,amsfonts}
\usepackage{algorithmic}
\usepackage{algorithm}
\usepackage{array}
\usepackage[caption=false,font=normalsize,labelfont=sf,textfont=sf]{subfig}
\usepackage{textcomp}
\usepackage{stfloats}
\usepackage{url}
\usepackage[hidelinks]{hyperref}
\usepackage{verbatim}
\usepackage{graphicx}
\usepackage{cite}
\usepackage{siunitx}

\begin{document}

\title{Adaptive Kinematic Modeling for Improved Hand Posture Estimates Using a Haptic Glove}

\author{\IEEEauthorblockN{Kathrin Krieger$\ast$, David P. Leins, Thorben Markmann, Robert Haschke, Jianxu Chen, Matthias Gunzer, and Helge Ritter}

        % <-this % stops a space
%\thanks{Manuscript received April 19, 2021; revised August 16, 2021.}
\thanks{The first three authors contributed equally to this work. The asterisk
indicates the corresponding author.}
\thanks{T. Markmann gratefully acknowledges funding from MKW NRW in the frame of the project SAIL (NW21-059A)}
\thanks{K. Krieger, D.P. Leins, R. Haschke, and H. Ritter are with the Neuroinformatics Group at CITEC, Bielefeld University}%
\thanks{T. Markmann is with the Machine Learning Group at CITEC, Bielefeld University}%
\thanks{K. Krieger, J. Chen, and M. Gunzer are with the Leibniz-Institut für Analytische Wissenschaften - ISAS - e.V}}%

%{Krieger \MakeLowercase{\textit{et al.}}: A Sample Article Using IEEEtran.cls for IEEE Journals}


\maketitle

\begin{abstract}
Most commercially available haptic gloves compromise the accuracy of hand-posture measurements in favor of a simpler design with fewer sensors. While inaccurate posture data is often sufficient for the task at hand in biomedical settings such as VR-therapy-aided rehabilitation, measurements should be as precise as possible to digitally recreate hand postures as accurately as possible. With these applications in mind, we have added extra sensors to the commercially available Dexmo haptic glove by Dexta Robotics and applied kinematic models of the haptic glove and the user's hand to improve the accuracy of hand-posture measurements. In this work, we describe the augmentations and the kinematic modeling approach. Additionally, we present and discuss an evaluation of hand posture measurements as a proof of concept.
\end{abstract}


\section{Introduction}
\label{sec:intro}
\IEEEPARstart{H}{aptic} gloves are wearable devices that provide force and/or tactile feedback to the fingertips and, optionally, the palm.
Further, they track the movement of the fingers and, in some cases, the palm with multiple degrees of freedom (DoF)~\cite{Wang2019HDF}.
In recent years, haptic gloves have become increasingly popular. 
Current literature envisions that such haptic gloves, if properly designed, will enable dexterous manual interaction with a virtual reality (VR) that feels realistic~\cite{Wang2018TWH}. 
However, haptic glove applications are not only limited to VR interaction.
Other examples range from robotic teleoperation~\cite{zhou2014rml,Kuling2020HFI} over surgical education~\cite{Abbas2020TRO} to behavioral analysis~\cite{wang2020digital}.
% example stroke rehabilitation with cyberglove, cyberforce and co~\cite{Alamri2008HVR,Alamri2007HEF,Kayyali2007occupational}

% examples research, 
% here reinforcement learning of participant performance in VR with Dexmo~\cite{heitkamp2022optimizing}

% application for mental health~\cite{wang2021supporting}
% product design application~\cite{malik2020nteract}

Haptic devices have also been applied for medical settings like motor rehabilitation~\cite{robotics10010040,Alamri2008HVR,Alamri2007HEF,Kayyali2007occupational} and medical training~\cite{Abate2014haptic,Abate2010virtual,Xue2020social}.
In these biomedical contexts, such devices are helpful because, in addition to measuring hand and finger movements, they also provide performance test scenarios by haptically representing virtual objects. Further, they allow the creation of applications hand-tailored to a patient's needs~\cite{Barros2014UOI,Alamri2008HVR}.

However, as Wang et al.~note, in practice, commercially available haptic gloves generally sacrifice the accuracy of motion tracking for a lower price and weight~\cite{Wang2018TWH}.
For example, the Dexmo glove\footnote{\url{https://www.dextarobotics.com/en-us}}~\cite{Gu2016DAI,DexmoPatent} only measures the overall flexion of each fingertip relative to its respective metacarpophalangeal joint (MCP) with a bend sensor, as well as the abduction of the MCP.
% However, these measurements do not allow inferring the angles of MCP, proximal interphalangeal (PIP), and distal interphalangeal (DIP) joints because for a single flexion value of a finger, the configuration space of MCP, PIP, and DIP does not offer a unique solution, i.e., the problem is underconstraint.
Still, a single bend value does not uniquely determine the flexion angles of all joints, i.e., MCP, proximal interphalangeal (PIP), and distal interphalangeal (DIP).
To circumvent this problem, the Dexmo software uses a linear regression model that infers PIP, DIP, and MCP angles based on the bend value of a finger~\cite{Gu2016DAI}. The drawback of this approach is that it constrains the measured position of individual fingertips to be on a circular trajectory between full extension and a power grasp.
This constraint limits the quality of force feedback, which depends on a high-quality fingertip pose estimation, and drastically reduces the number of natural movements the device can track.
For example, a precision pinch where PIP and DIP of the digit opposed to the thumb are fully extended, can not be captured.

\begin{figure}[t]
    \centering
    \includegraphics[width=\linewidth]{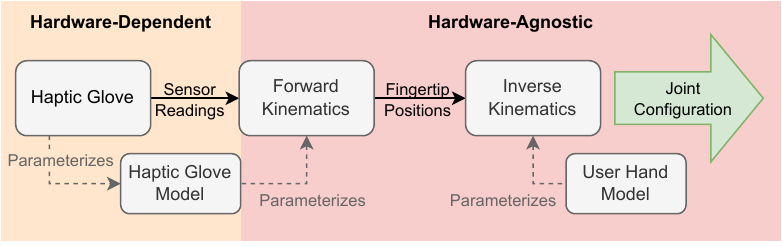}
    \caption{Data flow diagram of our model-based approach. The haptic gloves data is converted to fingertip positions by applying Forward Kinematics derived from a model of the glove. Based on a model of the user's hand, the application of Inverse Kinematics yields a valid joint configuration.}
    \label{fig:component_diagram}
\end{figure}

% Citation a la "For rehabilitation therapy movements should be as exactly replicated digitally" könnte hier hilfreich sein
Another challenge we identify in replicating hand postures digitally as accurately as possible is choosing the dimensions of the digital hand model. The lengths of the bones in the human hand dictate the positions of the joints and their influence on the fingertip positions relative to the palm. If the user's hand differs too much from the model, some postures might be physiologically impossible to reach with the digital hand.
To the best of our knowledge, no efforts have been made in the haptics community to develop an adaptive human hand model for haptic glove postures with a fast and reliable calibration method. Most gloves are under-determined but still compute the full hand joint states~\cite{caeiro_review}. Thus, they must resort to some variation of the mentioned coupling simplification of PIP and DIP to the MCP and use two extreme configurations to create a linear mapping between the joint of the real and digital hand that uses the digital model's full range of motion.

This work presents a modified version of the Dexmo haptic glove that allows us to estimate each finger's MCP, PIP, and DIP angles without constraining them to a predefined trajectory. We propose a framework that computes the hand joint configuration using kinematic modeling of both the glove and hand, along with sensor data.
As a part of this framework, we utilize the Hand Model Configuration Tool (HMCT)~\cite{kriegeropen} to measure the size of the user's hand and appropriately adapt the generic hand model employed for Inverse Kinematics.

The framework can be split into a hardware-dependent and a hardware-agnostic part (see Fig.~\ref{fig:component_diagram}). The hardware-dependent part consists of the glove itself and the software component tied to it that sends out its measurements, as well as a reliable model of the glove, its links, and its joints. Section~\ref{sec:augdex_fk} discusses this part of our framework applied to Dexmo in more detail. 
The hardware-agnostic part consists of (1) a Forward Kinematics module, which -- parameterized by the glove's exoskeleton model -- converts the incoming measurements into fingertip positions, and (2) an Inverse Kinematics module, parameterized by a configurable user hand model, that converts the fingertip positions into a hand joint configuration for downstream applications. This approach is detailed in section~\ref{sec:kinematic_models}.

\begin{figure}[tbp]
    \centering
    \includegraphics[width=0.75\linewidth]{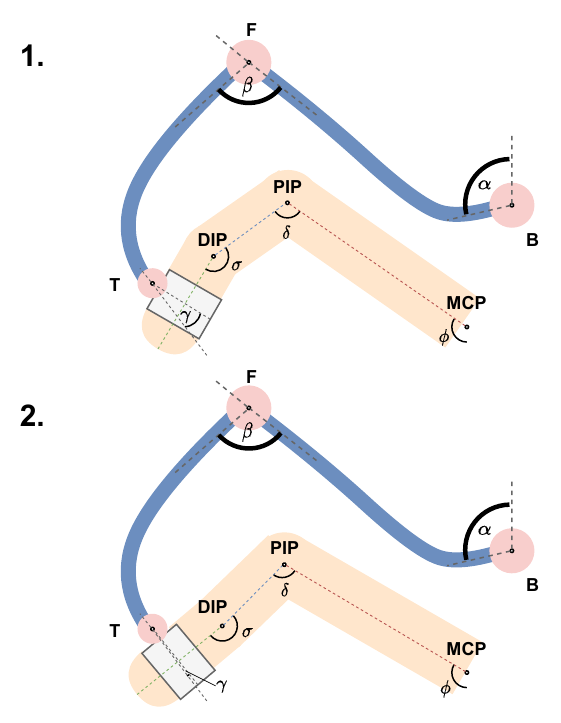}
    \caption{Concept of additionally measured angles for each finger. \textbf{B} denotes the bend sensor of Dexmo, which measures the flexion angle $\alpha$. \textbf{F} and \textbf{T} denote the newly added sensors to measure angles $\beta$ and $\gamma$ to compute the fingertip's position. Inverse Kinematics then solves for the finger-joint angles $\sigma$, $\delta$, and $\phi$. \textbf{1.} and \textbf{2.} display two distinct configurations with different configurations of DIP and PIP, but the same values for angles $\alpha$ and $\beta$, as an example of the under-determination problem, when only measuring $\alpha$.}
    \label{fig:augdex-concept}
\end{figure}

\section{Methods}

\subsection{Augmenting Dexmo}
% \cite{Friston2019PBC} they did very similar stuff to solve the issue with the under-determination, but position based with something on the other side of the hand. 
As already described in section~\ref{sec:intro}, Dexmo measures only a single bend value for the flexion of a finger, which measures the angle of the top bar ($\alpha$ in Fig.~\ref{fig:augdex-concept}), given a calibration that maps raw values to angles. Without further assumptions beyond the joint limits, the joint configuration can not be derived with that information alone. An example is shown in Fig.~\ref{fig:augdex-concept}, which depicts two different finger joint configurations with the same angles for $\alpha$ and $\beta$.
Given the length of the rods, adding an extra sensor to measure $\beta$ allows us to compute the position of the tip joint T. However, more information is needed to disambiguate the joint configuration. $\gamma$ must be known by adding a second sensor to determine the position and orientation of the T joint. 

\begin{figure}%[bp]
    \centering
    \includegraphics{figures/AugDex_explanation.jpg}\\
    \includegraphics[width=0.68\linewidth]{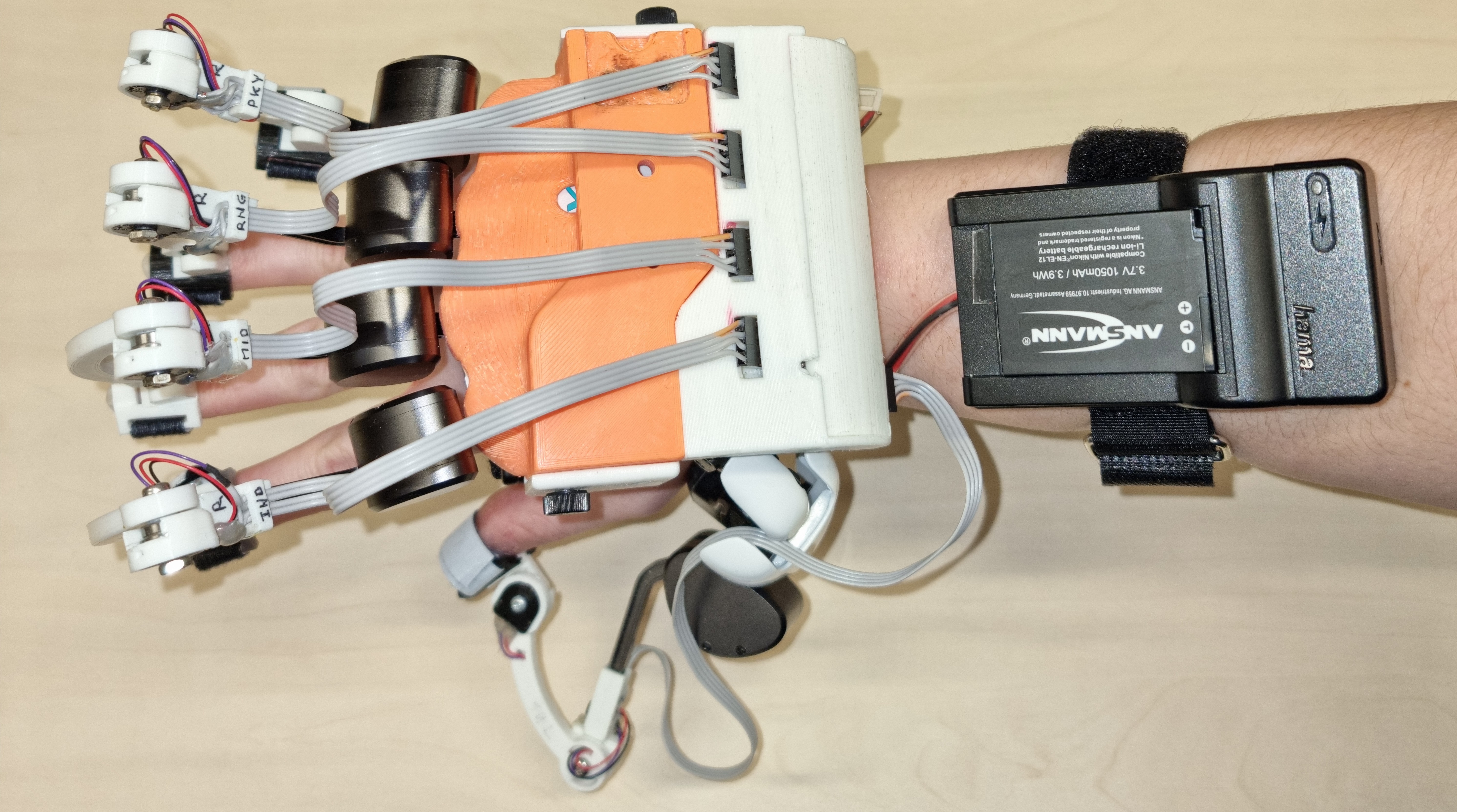}
    \caption{\textit{Top:} Labeled overview of the augmentation of Dexmo. \textit{Bottom:} Augmented Dexmo in action.}
    \label{fig:augdexmo_hardware}
\end{figure}

% @Thorben: Die Lösung der IK ist immer noch nicht eindeutig, oder?
% können wir irgendwie als kurzen Beweis die Anzahl möglicher Lösungen (mit nicht zu kleiner Auflösung) einer Beispielkonfiguration mit den default Limits fürs Handmodell berechnen für a. nur alpha is bekannt b. alpha und beta sind bekannt und c. alpha, beta und gamma sind bekannt?
% Fände ich etwas eleganter als nur stehen zu lassen, dass wir jetzt position und orientierung vom T joint messen können.
We replaced the Dexmo's rods with 3D-printed ones to measure these two additional angles.
They have cutouts with potentiometers inserted at joints T and F, i.e., the joints connecting the rods and the fingertip mount, as shown in Fig.~\ref{fig:augdexmo_hardware}.
These additional potentiometers connect to a Teensy board inside a 3D-printed housing attached to the Dexmo device.
The Teensy board reads the sensor values, broadcasts them at a configurable frame rate, and is powered by a 12V battery dorsally attached to the user's arm with a strap mount (see Fig.~\ref{fig:augdexmo_hardware}).

We further reference this extended version of the Dexmo exoskeleton glove as "Augmented Dexmo".

\subsection{Kinematic Models}
\label{sec:kinematic_models}
The approximation of the angles of MCP, PIP, and DIP joints is based on kinematic theory.
Kinematics is a field concerned with the motion of bodies, or links, through space and time. Joints are used to connect these links within kinematic models, which are tree structures with parents and children. A kinematic chain is a path from the root to an end effector. The Cartesian relation between a parent and a child is defined by a joint and a set of parameters such as the joint value $\theta$, a rotational axis, and a positional offset. These parameters usually describe the parent-child relation as a homogeneous transformation T.

Computing an end effector pose $\mathrm{T}_{\mathrm{EE}}(q)$ using a joint configuration $q = (q_1, ..., q_n)$ is a trivial task achieved by sequentially passing through the single joint transformations of the kinematic chain. This process is also known as Forward Kinematics (FK), and its solution is given by
\begin{equation}
    \mathrm{T}_0^{EE}(q) = f(q) = A_1(q_1)A_2(q_2)...A_n(q_n)
\end{equation}
where the $A_i$ are the homogeneous transformation matrices of individual joints, and $n$ is the number of joints in the kinematic chain.

Inverse Kinematics, on the other hand, solves the reverse problem of Forward Kinematics, i.e., determining the joint configuration leading to a given target end effector pose.
\begin{equation}
    q = f^{-1}(T_0^{EE})
\end{equation}
This problem is inherently more complex than FK since inverting the function $f$ is not trivial. Further, there could be no solution because the target pose is unreachable, e.g., due to joint constraints, or there may be an infinite number of solutions, e.g., if the kinematic chain has more DoF than the task itself.

Our implementations of FK and IK are based on MoveIt\cite{Coleman2017}, a framework for "motion planning, manipulation, 3D perception, Kinematics, control, and navigation" integrated into the Robot Operating System (ROS)\cite{ros}. MoveIt offers an API to solve FK on kinematic models and provides various IK solvers. Integrating ROS makes it possible to use the broad spectrum of ROS tooling, e.g., rviz, to visualize the haptic glove and the human hand model.

Overall, the Kinematics pipeline shown in Fig.~\ref{fig:component_diagram} consists of two main parts: an FK module to determine the fingertip positions using the measured joint values of the haptic glove's exoskeleton and an Inverse Kinematics module to compute the corresponding joint values of the human hand.

\begin{figure}[btp]
    \centering
    \includegraphics[width=.5\linewidth]{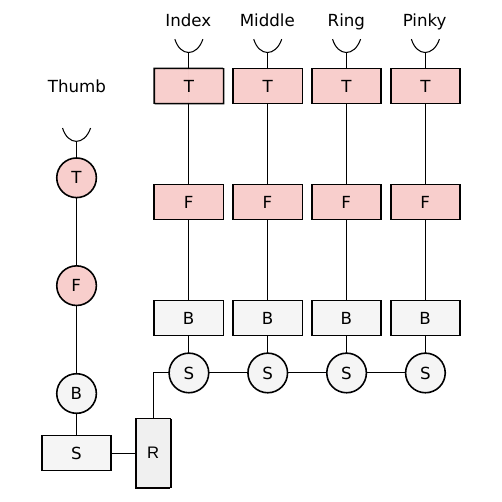}
    \caption{Schematic representation of the Augmented Dexmo model. Proximal, we used the original device~\cite{Gu2016DAI,DexmoLink} and the original model~\cite{DexmoPatent} (gray joints \textbf{R}otate, \textbf{S}plit and \textbf{B}end). Distal, we replaced the joints with our own augmented hardware (red joints \textbf{F}, \textbf{T}).}
    \label{fig:dexmo-model}
\end{figure}

\subsubsection{Augmented Dexmo -- Forward Kinematics}
\label{sec:augdex_fk}
As seen in Fig.~\ref{fig:component_diagram}, the FK module is hardware-agnostic, but it depends on a hardware-dependent kinematic model of the haptic glove and sensor readings from the actual glove. The kinematic glove model must include additional end effector links located at the position of the actual fingertips. In the case of Augmented Dexmo, these extra links are determined by a static relative transformation from the T-joint to the tip of the finger. We have settled on the naming convention "$\langle$finger name$\rangle$tip" to allow the FK module to identify these links. A schematic overview of our kinematic model of Augmented Dexmo is shown in Fig.~\ref{fig:dexmo-model}.

The other hardware-dependent part is the sensor reading of the haptic glove joints. The FK module expects sensor readings of the glove as tuples of joint values in radians and the name of the corresponding joint in the kinematic model. Since we read out the raw sensor values of Dexmo's joints, we have to determine a calibration of raw to angular values first. This step can be tedious and was done manually with the help of 3D-printed molds. It should be noted that this step is essential since calibration errors lead to wrong fingertip positions, which could result in inherently inaccurate approximations for the finger joint angles.\\

\begin{figure}[tbp]
	\centering
	\includegraphics[width=.9\linewidth]{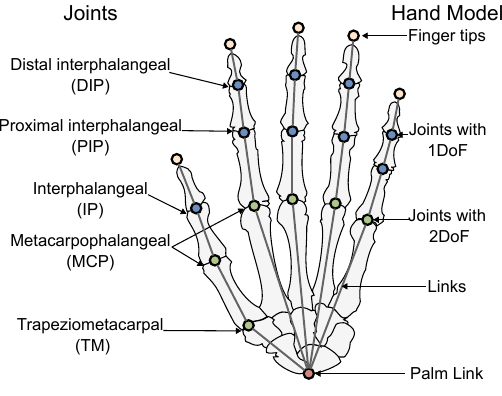}
	\caption{Dorsal, schematic representation of bones and joints in the right hand in extracts based on~\cite{Nanayakkara2017TRO,Nierop2008ANH,Chen2013CSF} and our simplified hand model.}
	\label{fig:hand-model}
\end{figure}

% \subsubsection{Hand Model -- Inverse Kinematics}
\subsubsection{Hand Model}
\label{sec:hand_ik}
The IK module and the hand model are completely hardware-agnostic, as shown in Fig.~\ref{fig:component_diagram}, i.e., they could be reused for any other haptic glove because they solely rely on the fingertip poses. 

The hand model is based on the model of the Pisa hand\cite{pisahand} and is publicly available\footnote{\url{https://github.com/DavidPL1/human_arm/tree/haptic-control}} in the URDF/SRDF format to be easily integrated into MoveIt. The kinematic structure of the hand model is very similar to that of a human hand shown in Fig.~\ref{fig:hand-model} but implements the 2 DoF MCP by two 1 DoF joints. Overall, there are five kinematic chains, one for each finger with 4 DoF or 5 DoF for the thumb chain, respectively, summing up to 21.
However, due to particular difficulties in aligning the thumbs of the Dexmo and hand model, the proposed framework omits the thumb entirely for now.

The hand model enables running Forward Kinematics to apply a set of joint angles to compute the fingertip positions and the backward path to compute a possible joint angle configuration from fingertip positions.
However, the finger segments' length and relative position considerably affect Forward and Inverse Kinematics. Thus, the model requires adaptation to fit the wearer's physique for more accurate angle computations. We use HMCT~\cite{kriegeropen} to measure the user hand quickly and reliably and generate a configuration for the hand model.
This user-specific configuration only has to be generated once, as the measured properties of the hand are not expected to change. 

\subsubsection{Inverse Kinematics}
\label{sec:ik_and_ref}
\begin{figure}[tbp]
	\centering
	\includegraphics[width=.85\linewidth]{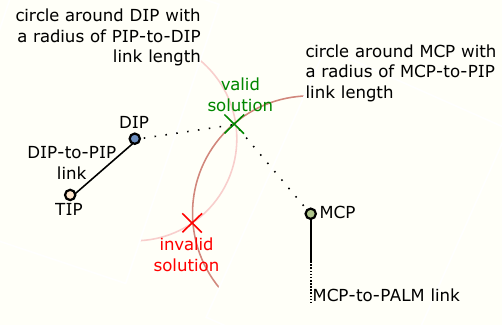}
	\caption{Schematic and geometric explanation of why the IK problem has a unique solution (if one exists at all). The given TIP pose and the DIP-to-TIP transform determine the DIP's pose. Further, the MCP pose is known from the PALM-to-MCP transform. Geometrically, two solutions exist given by the intersections of circles around the MCP and DIP that have the radius of the adjacent links. These intersections represent the PIP joint position. However, one of those solutions would yield a negative PIP joint angle and thus should be discarded. Hence, only one valid position for the PIP remains.}
	\label{fig:ik-has-one-solution}
\end{figure}

The IK module takes the calculated fingertip poses as input and solves the IK problem for the hand model. Our IK implementation is based on the MoveIt-included Orocos KDL\footnote{\url{https://github.com/orocos/orocos_kinematics_dynamics}} solver, an iterative Jacobian-based approach. The result is the joint configuration of the hand model to reach the finger target poses. If the IK solver cannot find an exact solution, the approximate solution generated by the iterative solver is returned. As demonstrated in Fig.~\ref{fig:ik-has-one-solution},  the IK solutions for the finger chains are unique if an exact solution is found.

When calculating an IK solution for the hand model based on the haptic glove's fingertip poses, the haptic glove space and the hand space should be properly aligned with a relative transformation. This alignment is determined during a calibration procedure, where a user wearing the glove places their index on a 3D-printed object such that both index and glove joint configurations are known, i.e., their end effector poses in world coordinates can both be computed by FK. Moving the glove space so that both poses are aligned in world coordinates ensures proper alignment between hand and glove models.

\subsection{Evaluation}
We conducted a study to validate the concept of kinematic modeling. Our aim was to demonstrate that this approach can provide accurate estimates of real finger joint angles.
Additionally, we compared the quality of our estimates with those generated by the Dexmo software, which we consider to be state-of-the-art.
%Further, we wanted to illustrate the impact of different factors on the quality of the outcome.\\

\subsubsection{Main Task}
\begin{figure}[tb]
	\centering
	\includegraphics[width=.7\linewidth]{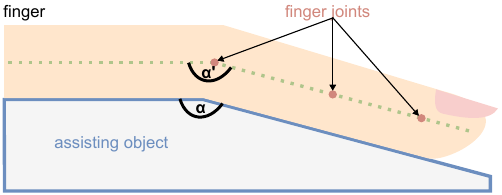}
	\caption{Illustration how the \textit{assisting objects} helped participants to perform specified finger postures.}
	\label{fig:augDexmo-eval-ideaObjects}
\end{figure}

To compare real finger joint angles with the estimated ones, we created a controlled environment where the angles were known. The setup involved the use of 3D-printed objects enclosing a target angle, $\alpha$, between two flat surfaces, as illustrated in Fig.~\ref{fig:augDexmo-eval-ideaObjects}. Participants were required to bend a joint at approximately $\alpha$ while keeping the other joints straight by aligning the edge with the joint and pressing their finger against the surfaces.

\subsubsection{Procedure}
Initially, participants signed an informed consent and filled out a short general questionnaire to characterize the participant's sample.
A hand model configuration was created using the HMCT~\cite{kriegeropen} for the participants' right hand.
Participants then continued to put on the Augmented Dexmo on the right hand; the offset between the haptic glove and the hand was calibrated with the procedure described in sec.~\ref{sec:ik_and_ref}, and the experimental trials were executed.
A calibration profile was created with the original Dexmo software and calibration procedure to allow for comparison with Dexmo's original estimates.
In the end, participants received fruits or sweets to compensate for their time.

\subsubsection{Experimental Design}
A prior pilot evaluation with two other participants showed no qualitative differences between the estimates of the individual fingers.
Consequently, we could limit the experiments to the index finger's bending joints MCP, PIP, and DIP to keep the experimental design condensed.
We chose two far-apart target angles to represent the full range of motion best.
For the first angle, we opted for \ang{20}, considering the glove's lower joint limits with an added safety margin.
To ensure comfort, we set the second target angle to \ang{45} considering the finger joints' upper limits~\cite{Mallon1991DRO} while also adding a margin.
Accordingly, the study was a 3 (specified joint) x 2 (target angle) within-subjects design.
In order to also measure repeatability, each of the six experimental conditions was repeated six times, resulting in 36 trials for each participant.
To avoid biases, the 36 trials were organized in six successive blocks, each one containing all six experimental conditions once in an individually randomized order.

\subsubsection{Participants}
We performed a power analysis to determine the number of participants required to find a qualitative difference between the joint angle estimates, the kinematic modeling, and the original estimates.
Assuming to find a noteworthy difference $d=0.8$ between the paired estimate samples with $\alpha=0.05$ and $1-\beta=0.80$, we concluded our evaluation to require twelve participants.

Thus, 12 participants (9 male, 3 female) joined the study, of which 11 were right-handed and one was left-handed.
Their ages were between 24 and 61 years, with a median age of 34.
None of them reported any haptic or motor impairment.
Their hand length, from the wrist to the index fingertip, ranged from 15.2\,cm to 19.2\,cm ($\mu$ 17.2\,cm).
The ethics committee of Bielefeld University has approved the study protocol under number 2022-282.

\subsubsection{Data Recording and Analysis}
For each trial, we saved the specified joint to be bent (\textit{bent joint}) and its target angle.
Moreover, we recorded a five-second trajectory of the Augmented Dexmo's raw and calibrated B, F, and T joints. 
Based on these, our kinematic modeling framework and the Dexmo software calculated the estimates for all finger joint angles, respectively. 
Afterward, we computed the \textit{mean angle} of each joint during the five-second trajectories for both estimates.
To gather insights into the quality of the estimates, we investigated four different metrics, which are based on calculating the angular distance $\Phi$ between the estimate and target angle.
The first metric $\Phi_{\mathrm{bent joint}}$ provides a global view of how well the frameworks performed in each experimental condition. 
For this, we calculated the angular distance between the estimated angle of the \textit{bent joint} and the target angle for each trial.
In order to examine the quality of the joint angle estimation in more detail, the three other metrics $\Phi_{\mathrm{MCP}}$, $\Phi_{\mathrm{PIP}}$, and $\Phi_{\mathrm{DIP}}$ explored the MCP, PIP and DIP individually.
Thus, for each metric and for each trial, we computed the angular distance between the expected angle and the joint angle estimate.
Here, an angle of \ang{0} is assumed as the expected angle for experimental conditions in which the joint should not be bent.
For each participant, we computed a mean for each metric.
We used two-sided paired t-tests with $\alpha=0.05$ to compare the metrics derived from the estimates of the kinematic modeling framework and the Dexmo software.

\section{Results}
Table~\ref{tab:stats} shows the four metrics' means and standard deviations over all participants for the kinematic modeling framework and the Dexmo software and the outcome of the t-tests.
Comparing the metric $\Phi_{\mathrm{bent joint}}$ between the kinematic modeling and the Dexmo software, we notice considerable differences.
The kinematic modeling approach, on average, yields significantly smaller angular distances with $p = 0.002$. While $\Phi_{\mathrm{MCP}}$ and $\Phi_{\mathrm{DIP}}$ show slightly higher angular distances, proven insignificant by the t-test, $\Phi_{\mathrm{PIP}}$ displays a significant improvement with $p < 0.001$ compared to the standard Dexmo software.

\begin{table}[tbp]
\centering
\begin{tabular}{l c c c c} 
 & $\Phi_{\mathrm{bent joint}}$ & $\Phi_{\mathrm{MCP}}$ & $\Phi_{\mathrm{PIP}}$ & $\Phi_{\mathrm{DIP}}$\\ 
 \hline
 $M_{\mathrm{kinematic}}$&15.45 & 12.12 & 13.37 & 15.34\\ 
 $M_{\mathrm{Dexmo}}$&20.04 & 10.68 & 18.02 & 13.82 \\
 $SD_{\mathrm{kinematic}}$&3.36 & 2.91 & 2.93 & 6.19 \\
 $SD_{\mathrm{Dexmo}}$&2.12 & 2.20 & 1.04 & 1.61 \\
 $t(11)$&-4.06 & 2.19 & -4.92 & 0.75 \\ 
 $p$&0.002&0.051&$<$0.001& 0.468\\
 \hline
\end{tabular}
\caption{Statistics of the four metrics describing angular distances between estimates and target angles.}
\label{tab:stats}
\end{table}

\begin{figure}[bp]
	\centering
 \includegraphics[width=.95\linewidth]{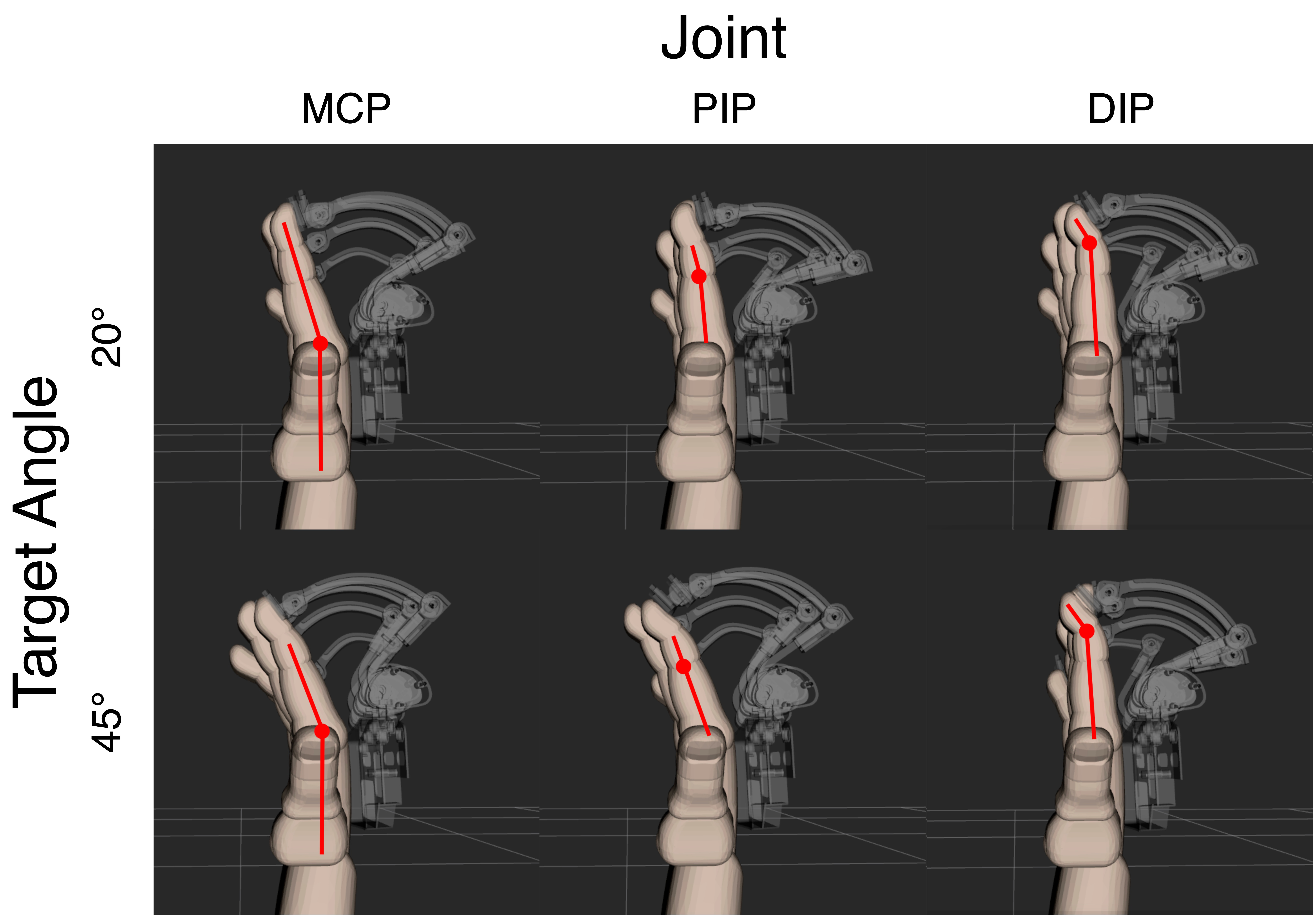}
    \caption{Exemplary presentation of Augmented Dexmo and hand model visualization in rviz for the six experimental conditions.}
	\label{fig:qualitative-results}
\end{figure}

For each condition, we created screenshots of Augmented Dexmo and hand model visualizations in rviz.
These screenshots are presented in Fig. ~\ref{fig:qualitative-results} and show, in most cases, that the estimation plausibly approximates the \textit{bent joint}.
Based on visual inspection, it appears that the DIP joint is more susceptible to incorrect bending than the PIP and MCP joints.

\section{Discussion}
The study results indicate that our approach can estimate independent joint movements more accurately than the standard Dexmo software.
However, the quantitative evaluation revealed a mean angular distance of approximately \ang{15} between the estimated and the target joint angles, with the DIP joint displaying the lowest accuracy. The IK predominantly alters the DIP angle to balance offsets in the kinematic chains. This is a clear drawback of not considering the natural coupling of the PIP and DIP joints~\cite{Mallon1991DRO}. We do not achieve any significant improvement or decline in the estimation for the MCP, which is also Dexmo's most accurately estimated joint. Lastly, the PIP shows an average improvement of approximately \ang{5} compared to Dexmo, which is a noticeable amount considering the investigated range of motion. Since MCP and PIP with considerably longer link lengths than the DIP contribute stronger in the final pose of a finger, we consider our approach a successful step towards using haptic gloves for precise posture measurements or natural interaction in VR. 

In theory, it should be possible to calculate the correct angles with the kinematic models. However, In practice, the accuracy is limited to a) accurate models of glove and hand, where small offsets manifest in systematical, propagating errors of the angle estimation, and b) the consistency of the glove's offset calibration. In trials, we noticed the latter to be challenging to maintain because, due to Dexmo's mount design of using only a single velcro strap at the palm center and the heavy design of the glove, it is prone to slip even during slight movements.

For future work, we aim to extend the IK with a constrained solver that integrates the coupling between PIP and DIP, which should reduce the error rate of the DIP. Further, we want to extend our model to produce reasonable estimates for the thumb.
Finally, we plan to extend our framework to other gloves, i.e., SenseGlove DK1~\cite{SenseGloveHandbuch}, to leverage the hardware-agnostic design and provide a versatile tool for the haptics community.
% As additional glove, the , as it already measures the same joint angles as our Augmented Dexmo device.
% Finally, we want to show that using our framework for improved posture measurement helps people to naturally interact with a virtual environment, compared to using that haptic gloves only.

In conclusion, we are working towards improving haptic gloves with an adaptive kinematic modeling framework to estimate precise hand posture measurements for natural manual data recording, and interaction in VR. 
The main contribution of this paper is the proposed framework.
We have shown results on how the framework improves on the state-of-the-art natural hand posture estimation and discussed how we aim further to reduce the gap between real and estimated joint angles. 
Together with this paper, we publish the open-source hardware-agnostic part of our framework.
% , which consists of two parts. 
% The first part is hardware-dependent and determines the haptic glove's end effector position and orientation.
% The second part is hardware-agnostic, where we use a calibrated hand model and the known fingertip position and orientation to determine the hand posture with an IK algorithm.

\section*{Aknowledgements}
We would like to thank Guillaume Walck for the many hours of discussing the concept, building and programming the augmentation hardware that measures the T and F joints, and working on a server software.

\bibliographystyle{IEEEtran}
\bibliography{bibfile}

\end{document}